\documentclass[aps,prl,twocolumn,letterpaper,superscriptaddress,showpacs]{revtex4}
\usepackage[colorlinks,linkcolor=blue,anchorcolor=blue, citecolor=blue,urlcolor=blue,]{hyperref}
\usepackage{graphicx}
\usepackage{CJK}
\usepackage{verbatim}
\usepackage{physics}
\usepackage{color}
\usepackage{mathptmx}
\begin{document}

\begin{CJK*}{UTF8}{bsmi}

\title{Chemical disorder induced electronic orders in correlated metals}

\author{Jinning Hou (\CJKfamily{gbsn}侯晋宁)}
\affiliation{Tsung-Dao Lee Institute \& School of Physics and Astronomy, Shanghai Jiao Tong University, Shanghai 200240, China}

\author{Yuting Tan (\CJKfamily{gbsn}谭宇婷)}
\affiliation{Department of Physics and National High Magnetic Field
Laboratory, Florida State University, Tallahassee, FL 32306, USA}

\author{Wei Ku (\CJKfamily{bsmi}顧威)}
\altaffiliation{corresponding email: weiku@sjtu.edu.cn}
\affiliation{Tsung-Dao Lee Institute \& School of Physics and Astronomy, Shanghai Jiao Tong University, Shanghai 200240, China}
\affiliation{Ministry of Education Key Laboratory of Artificial Structures and Quantum Control, Shanghai 200240, China}
\affiliation{Shanghai Branch, Hefei National Laboratory, Shanghai 201315, China}

\date{\today}

\begin{abstract}
In strongly correlated metals, long-range magnetic order is sometimes found only upon introduction of a minute amount of \textit{disordered} non-magnetic impurities to the unordered clean samples.
To explain such anti-intuitive behavior, we propose a scenario of inducing electronic (magnetic, orbital, or charge) order via chemical disorder in systems with coexisting local moments and itinerant carriers. 
By disrupting the damaging long-range quantum fluctuation originating from the itinerant carriers, the electronic order preferred by the local moment can be re-established.
We demonstrate this mechanism using a realistic spin-fermion model and show that the magnetic order can indeed be recovered as a result of enhanced disorder once the length scale of phase coherence of the itinerant carriers becomes shorter than a critical value.
The proposed simple idea has a general applicability to strongly correlated metals, and it showcases the rich physics resulting from interplay between mechanisms of multiple length scales.
\end{abstract}

\maketitle
\end{CJK*}

Typically, random disorder is expected to suppress long-range orders in materials, especially those with a characteristic length scale such as antiferromagnetic order, antiferro-orbital order, or charge density order.
This is in part because of the damage to quantum phase coherence resulting from the inhomogeneity in density, in addition to the direct disruption of the preferred spatial periodicity of the long-range order.
Indeed, in dirtier samples with more impurities, one usually observes weaker magnetic~\cite{Shen2017,Cheong1991,Vajk2002,Delannoy2009}, superconducting~\cite{schneider2012,mackenzie1998,fujita2005} and charge~\cite{sham1976effect,Straquadine2019,Fang2019,Fang2020} orders.
Correspondingly, one often intuitively seeks cleaner and more uniform samples for stronger long-range orders.


However, some exceptional cases exist in which long-range order, for example magnetic order, can emerge from the introduction of disorders, such as non-magnetic impurities.
A well-known example is the emergence of antiferromagnetic (AFM) order in Sr$_2$RuO$_4$~\cite{Braden2002,Minakata2001,Ishida2003,Pucher2002} when a minute amount ($\sim 3\%$) of Ru$^{4+}$ are substituted by non-magnetic Ti$^{4+}$ ions. 
Similarly, iron-based superconductor LaFePO also develops antiferromagnetism upon As substitution of P~\cite{Kitagawa2014,Lai2014,Mukuda2014}.
The indications that AFM order could emerge from unordered systems also have been found in hole-doped cuprates via Zn substitution of Cu, as measured by muon spin resonance ($\mu$SR)~\cite{Watanabe2000} and neutron scattering experiments~\cite{Hirota1997,Kimura2003,Suchaneck2010}. 
Such an anti-intuitive behavior appears to contradict the above fundamental consideration of quantum phase coherence, and thus poses a great challenge to our generic basic understanding.

Theoretically, in a strongly correlated and highly polarizable environment, it is natural to expect the development of local effective moments around even non-magnetic impurities~\cite{Wessel2001,Yasuda2001,Bobroff2009}.
Such an effective moment surely would have a large impact on the local correlation, such as modifying its temporal fluctuation or inducing a spatial standing-wave pattern through reflection against impurities~\cite{Martiny2019,Gastiasoro2015,Zinkl2021,song2020}.
Nonetheless, since these effects are primarily local in nature and centered around \textit{random} location of the impurities, it is unlikely that they can provide positive contributions to the formation of long-range order, especially those with a characteristic spatial period, such as an antiferromagnetic order. 
Therefore, a generally applicable mechanism for the observed seemingly anti-intuitive behavior is desperately needed for such a long-standing puzzle.

Here, we propose a generic scenario of inducing electronic order via a small amount of chemical disorder in strongly correlated metals.
Accepting that most of unordered correlated \textit{metals} only fail to order due to the influence of itinerant carriers~\cite{Tan2019,tam2015}, we suggest that chemical impurities can suppress the damaging carrier-induced long-range quantum fluctuation and in turn allow the local moments to order.
We demonstrate this generic mechanism using a realistic spin-fermion model derived from FeSe as a prototypical case with a \textit{failed} antiferromagnetic (AFM) order~\cite{Tan2019}.
Using the linear response as a measure of the stability of the AFM ordered state, we find that with a stronger disorder the long-range magnetic order indeed establishes.
Further analysis indicates that the main physical effect of impurity scattering is equivalent to shortening the length scale of carrier-induced quantum fluctuation, such that the correlation of local moments is no longer overwhelmed at long range~\cite{Tan2019}.
Our study demonstrates a typical example of the rich interplay between mechanisms of multiple length scales present in most strongly correlated metals, to which our proposed simple idea can be applied in general.

\begin{figure}
	 \setlength{\belowcaptionskip}{-0.5cm}
	\vspace{-0.1cm}
	\begin{center}
	    \includegraphics[width=8.5cm]{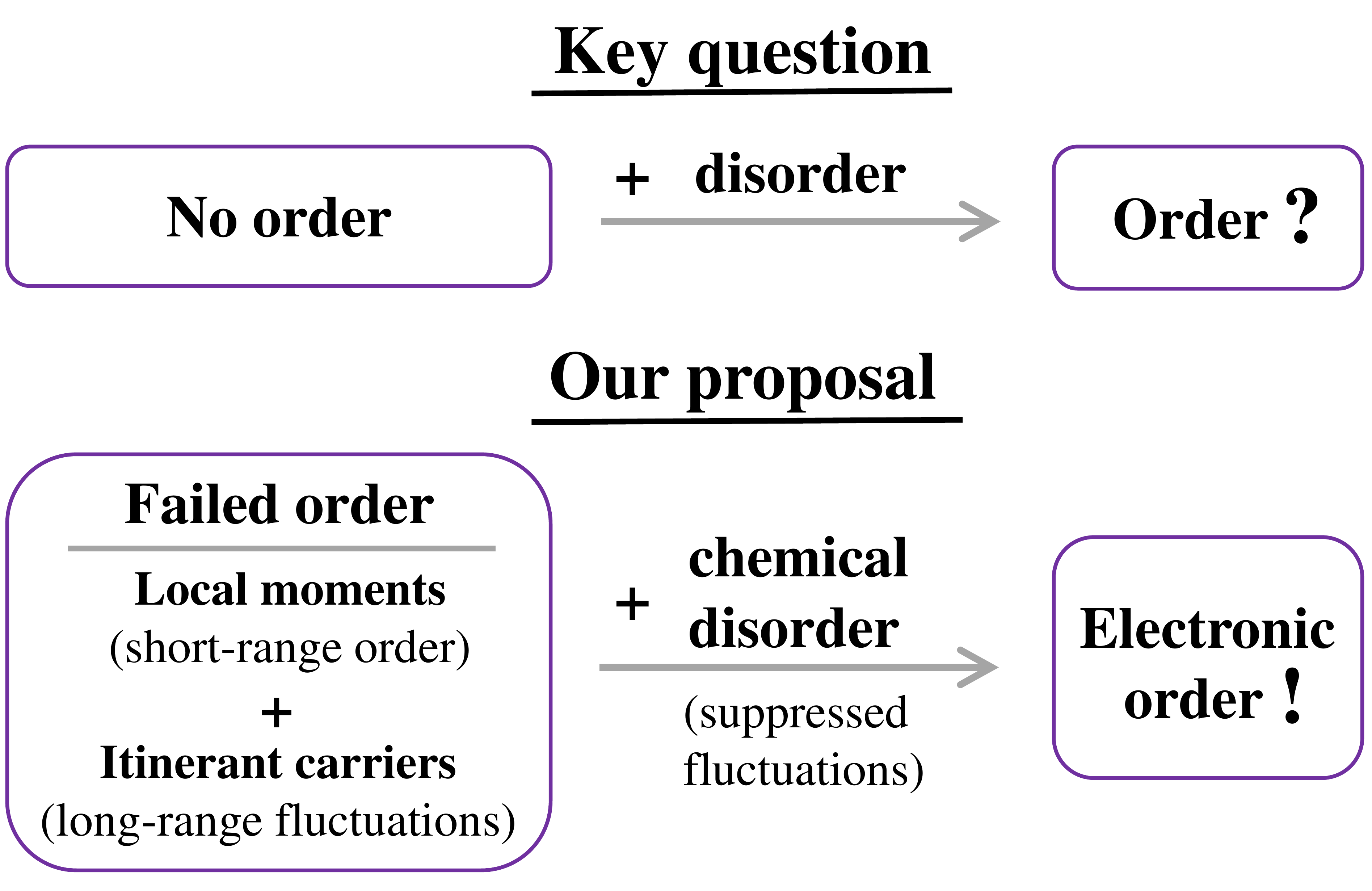}
	\caption{Key question of the study: How can chemical disorder induce electronic orders in materials? Our proposal: Charge disorder can suppress the damaging long-range fluctuation of a failed order and in turn allow the intrinsic electronic order to emerge.
	}
	\label{fig1}
	\end{center}
\end{figure}

Figure~\ref{fig1} illustrates our proposed scenario to resolve the long-standing puzzle of electronic ordering upon the introduction of chemical disorder in correlated metals.
The key theoretical question here is how disorder, a generic source of incoherence, can induce a coherent long-range order.
Our proposal is based on a ``failed order" scenario~\cite{Tan2019} in which the long-range order preferred by the correlation between local moments is disrupted by the long-range quantum fluctuation induced by itinerant carriers~\cite{tam2015}.
Such quantum fluctuation can be quite effective in general since in contrast to the exponential decay of the order-related correlation in three dimensions, the carrier-induced fluctuation has generic power-law decay, due to the discontinuity at the Fermi surface of the fermionic carriers~\cite{jagannathan1988,bulaevskii1986rkky,sobota2007rkky}.
We therefore propose that by restricting the fluctuation to a short enough finite length scale, the presence of disorder can play a positive role in promoting the long-range order of the local moments.

Below we proceed to demonstrate this generic mechanism using a realistic spin-fermion model.
We first integrate out the influence of the itinerant carriers to second order, which associates their long-range fluctuation with the effective interaction between the local moments.
We then demonstrate the system's ``failed order'' nature using the linear response of the ordered state as a measure of its instability.
After that, we simulate the disorder effect numerically and confirm the establishment of long-range order.
Finally, we analyze the various emergent length scales in our result and provide an intuitive microscopic picture for the leading physics.

As a generic example, consider a realistic spin-fermion model consisting of coupled local moments affected by itinerant carriers~\cite{yin2010,tam2015,Tan2019,lv2010,kou2009,liang2012}:
\begin{equation}
    \label{spin_fermion}
    \begin{split}
        H &= \sum_{i \neq i^\prime} J_{ii^{\prime}} \mathbf{S}_{i} \cdot \mathbf{S}_{i^\prime} \\
        &- J_{\mathrm{H}}\sum_{im\nu\nu^\prime}\mathbf{S}_{i} \cdot c^{\dagger}_{im\nu} \mathbf{\sigma}_{\nu\nu^\prime}
        c_{im \nu^\prime} \\
        &+\sum_{jj^\prime mm^\prime \nu}t_{jmj^\prime m^{\prime}}c^{\dagger}_{jm\nu}c_{j^\prime m^\prime \nu},
    \end{split}
\end{equation}
where the local moments $\mathbf{S}_{i}$ at site $i$ and $i^\prime$ couple via $J_{ii^{\prime}}$ such that a magnetic stripe ($\pi,0$) order is preferred by the local moments~\cite{Tan2019}.
The non-trivial physics emerges when these local moments couple ferromagnetically to the itinerant carriers $c_{im\nu}^\dagger$ of orbital $m$ and spin $\nu$ at the same site $i$ via coupling constant $J_{\mathrm{H}}$, where $\mathbf{\sigma}_{\nu\nu^\prime}$ are the Pauli matrices.
This is because the itinerant carriers can propagate between sites with kinetic parameter $t_{jnj^{\prime}n^{\prime}}$ and are thus able to mediate an effective long-range interaction~\cite{jagannathan1988,bulaevskii1986rkky,sobota2007rkky} between the local moments at longer time scale (or lower energy) relevant to the slower spin dynamics.
Note that we consider a general case in which the fermion orbitals at sites $j$ can reside at the same site $i$ as the local moments or those without (such as ligand sites).

This emergent interaction can be obtained by integrating out the faster itinerant electron degrees of freedom.
For simplicity, we stick to the weak coupling regime where $J_{\mathrm{H}}$ can be considered a perturbation that renormalizes~\cite{Tan2019,tam2015,lv2010} the linear spin-wave theory~\cite{Anderson1952,Supplementary} corresponding to the preferred long-range order.
Represented in the second quantized magnon creation operator $a_{i}^{\dagger}$ associated with the Holstein-Primakoff transformation~\cite{HPtransformation1940}, the resulting spin-wave Hamiltonian reads:
\begin{equation}
\begin{split}
    \label{H_J_real_space}
    H^{\mathrm{SW}} &=\sum_{i} \tilde{K}_{i} a_{i}^{\dagger}a_{i} \\
    &+ \sum^\mathrm{FM}_{i\neq i'} \tilde{J}_{ii'}(a_{i}^{\dagger}a_{i'}+a_{i}a^{\dagger}_{i'})+\sum^\mathrm{AF}_{i\neq i''}\tilde{J}_{ii''}(a_{i}^{\dagger}a_{i''}^{\dagger}+a_{i}a_{i''}),
\end{split}
\end{equation}
where $\tilde{K}_{i} = 2\sum^\mathrm{AF}_{i''}\tilde{J}_{ii''}-2\sum^\mathrm{FM}_{i'}\tilde{J}_{ii'}$ ensures the preservation of the Goldstone mode of the ordered system.
Here the summation is split into those between the parallel (FM) and anti-parallel (AF) pairs of spins, with their coupling being renormalized by $\tilde{J}_{ii'}=J_{ii'}+A_{ii'}$ and $\tilde{J}_{ii''}=J_{ii''}+B_{ii''}$, respectively.
Represented in momentum $q$ space,
\begin{equation}
\label{A2}
\begin{split}
    A(q) = &\frac{J_{\mathrm{H}}^{2}}{2S}\sum_{kll^{\prime}}\frac{(f_{l}(k)-f_{l^{\prime}}(k+q))(E_l(k)-E_{l^{\prime}}(k+q))}{(E_{l}(k)-E_{l^{\prime}}(k+q))^2+\delta^2}\\
    &\times \left|\sum_{m}U^{l^{\prime}\star}_{m\downarrow}(k+q)U^{l}_{m\uparrow}(k)\right|^{2}\mathrm{,\ and}\\
    B(q) = &\frac{J_{\mathrm{H}}^{2}}{2S}\sum_{kll^{\prime}}\frac{(f_l(k)-f_{l^{\prime}}(k+q))(E_{l}(k)-E_{l^{\prime}}(k+q))}{(E_{l}(k)-E_{l^{\prime}}(k+q))^2+\delta^2}\\
    &\times \sum_{mm^{\prime}}U^{l^{\prime}\star}_{m\downarrow}(k+q)U^{l}_{m\uparrow}(k)U^{l\star}_{m^{\prime}\downarrow}(k)U^{l^{\prime}}_{m^{\prime} \uparrow}(k+q),
\end{split}
\end{equation}
where $E_l(k)$ denotes the eigenvalues with momentum $k$ and band index $l$ (that absorbs the spin index as well) and $U_{m\nu}^l(k)$ denotes the eigenvectors in the basis of orbital $m$ with spin $\nu=\uparrow$ or $\downarrow$.
$f_l(k)=\frac{1}{1+e^{\beta(E_l(k)-\mu)}}$ is the standard Fermi-Dirac distribution function for a given chemical potential $\mu$, and $S$ the effective magnitude of the local moments.
The typical numerical broadening of $\delta=0^+$ is not necessary here since we are only interested in the zero-frequency limit of the renormalization.

We now seek a ``failed order'' state as the unordered state prior to the introduction of chemical disorder.
It was recently suggested~\cite{Tan2019} that the semi-metallic FeSe is such a failed ordered system whose AFM order only appears under external pressure greater than 1GP when the carrier density decreases.
In essence, the reduction of carrier density weakens the carrier-induced long-range fluctuation and in turn allows the long-range AFM order of the local moments to emerge.
The fact that the failed order state can be overcome by mere 1GPa of pressure implies that the long-range fluctuation is close to being overcome by the ordering, making it an ideal model system for our demonstration.
We thus take the parameters of Eq.(\ref{spin_fermion}) from the previous density-function based study~\cite{Tan2019}, which incorporates $t_{jmj^\prime m^{\prime}}$ of five $d$-orbitals and three $p$-orbitals, $J_{\mathrm{H}}=0.8$ eV, $S=1.7$, and $J$=19meV and 12meV for the nearest and next nearest neighbors, respectively.
A discrete $500\times500$ momentum mesh and a 10meV thermal broadening are used to ensure a good convergence.

\begin{figure}
	 \setlength{\belowcaptionskip}{-0.5cm}
	\vspace{-0.1cm}
	\begin{center}
	    \includegraphics[width=8cm]{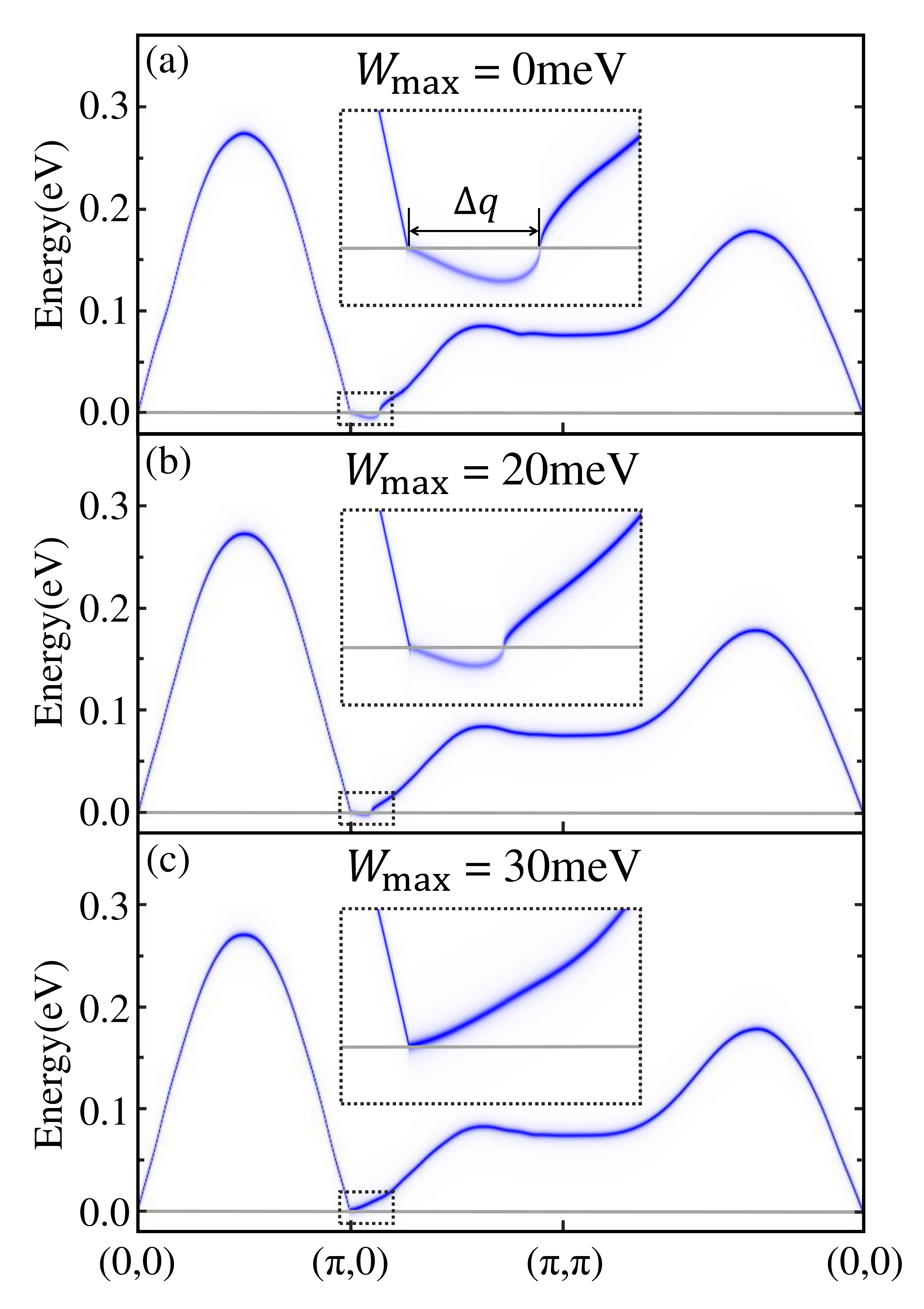}
	\caption{Emergence of long-range magnetic order via the introduction of disorder to a failed ordered metal with large magnetic moments. Magnetic susceptibility is shown here as a measure of the stability of the preferred ordered state. (a) In the clean system ($W_{\mathrm{max}}=0$) imaginary (shown as negative) frequency appears near ($\pi$,0) as magnified in the inset, indicating the ($\pi$,0) AFM order is unstable due to carrier-induced fluctuation at long range beyond $2\pi/\Delta q$. (b) Introduction of disorder with $W_\mathrm{max}=20$meV weakens the fluctuation (corresponding to a smaller $\Delta q$), allowing the correlation to extend to a longer $2\pi/\Delta q$ range. (c) By $W_\mathrm{max}=30$meV, all excitations energies become positive ($\Delta q=0$), indicating the correlation is now long-range and the preferred ($\pi$,0) magnetic order is a stable phase. That is, the electronic long-range order is induced by the introduction of chemical disorder.
	}
	\label{fig2}
	\end{center}
\end{figure}

Let's first verify the ``failed order'' state prior to the introduction of chemical disorder, by examining the stability of the ordered state via its linear response.
Figure~\ref{fig2}(a) and its inset show that in the absence of disorder, the obtained spin wave energy-momentum dispersion displays no positive-energy excitation in the vicinity of ($\pi$,0).
Such lack of positive-energy excitation in the linear response is a direct indication that the assumed AFM ordered state is unstable, in this case due to the carrier-induced long-range fluctuation that overwhelms the correlation at length scale longer than $2\pi/\Delta q$.
In other words, we have verified that our starting point is indeed a failed order state, in which the local moments are unable to establish long-range order even at the zero temperature limit.


\begin{figure}
	 \setlength{\belowcaptionskip}{-0.5cm}
	\vspace{-0.1cm}
	\begin{center}
	    \includegraphics[width=6.6cm]{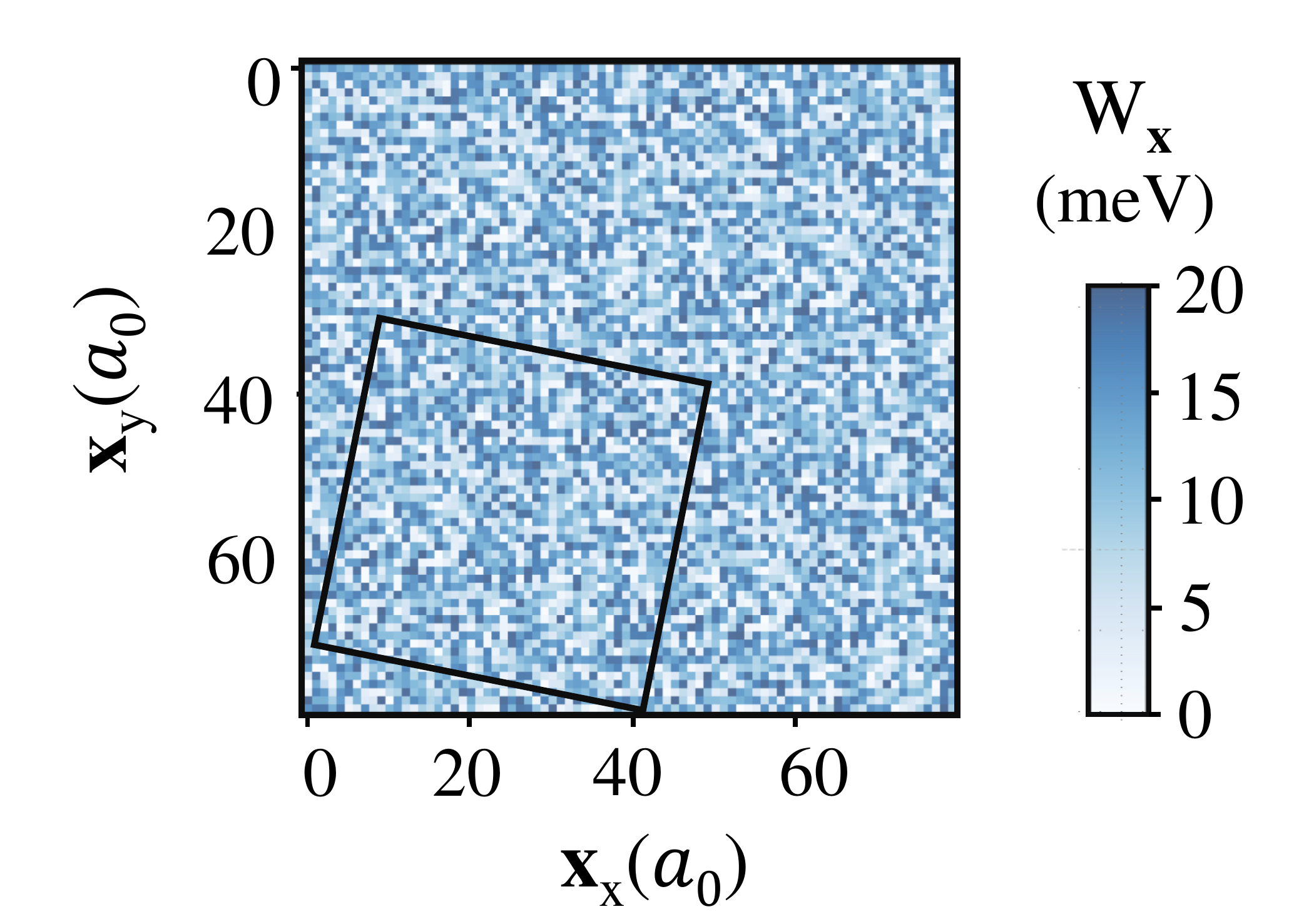}
	\caption{An example of disorder configurations containing 1664 sites with $W(\mathbf{x})$ randomly sampled from a uniform distribution within [0,$W_{\mathrm{max}}$].
    The orientation of the periodicity of the disorder configuration (denoted by the stilted square) is randomly chosen for each configuration so that the associated artifacts can be suppressed in the ensemble average.
	}
	\label{fig_disorder}
	\end{center}
\end{figure}

We now proceed to include the effect of chemical disorder-induced scattering on the itinerant carriers and investigate its effect on the long-range order.
Specifically, we aim to calculate the linear response of the long-range ordered state by ensemble-averaging over a large number of chemically disordered configurations.
It is well-established~\cite{jagannathan1988,bulaevskii1986rkky,sobota2007rkky} that the main effect of disorder on the magnetic quantum fluctuation of itinerant carriers is to introduce incoherent phase shifts along its propagation without affecting its power-law spatial decaying profile.
We therefore approximate the incoherent phase shifts in the fluctuation within each configuration according to~\cite{bulaevskii1986rkky}
\begin{equation}
\label{J_phi_dis}
    \tilde{J}_{ii^\prime} \longrightarrow \tilde{J}_{ii^\prime}\cos{\phi_{ii^\prime}},
\end{equation}
where
\begin{equation}
\label{phi_dis}
    \phi_{ii^\prime} = \frac{2}{\hbar v_{F}} \int_{\mathbf{x}_{i^\prime}}^{\mathbf{x}_i} ds W(\textbf{x}),
\end{equation}
accumulates phase shift from scattering against spatially random potential $W(\textbf{x})$ along a straight path from position $\textbf{x}_{i^\prime}$ of site $i^\prime$ to position $\textbf{x}_i$ of site $i$.
(see Supplementary~\cite{Supplementary} for detail on discretization of the disorder strength and its path integration.)
The strength of the disorder potential $W(\textbf{x})$ is randomly sampled from a uniform distribution between 0 and $W_\mathrm{max}$.
We apply Eq.(\ref{J_phi_dis}) to disorder configurations with large systems (typically containing around 1600 sites) of various shapes and orientations in the simulation~\cite{Tom2014,Tom2013,Tom2012,Tom2011}.
(See Fig.~\ref{fig_disorder} for an example.)
For each configuration containing different phase shifted $\tilde{J}_{ii^\prime}$ for each pair of $i$ and $i^\prime$ (Eq.~\ref{phi_dis}), the magnon spectral function is then calculated via numerical bosonic Bogoliubov diagonalization~\cite{Tsallis1978,Supplementary} of $H^\mathrm{SW}$ followed by the unfolding procedure~\cite{Wei_unfolding_2010} before being averaged over the ensemble.

Figure~\ref{fig2}(b) and (c) gives the resulting magnon energy-momentum dispersion under increasing disorder strength.
Since the main effect of disorder is through the phase shift of the carrier-induced long-range fluctuation, the physical broadening~\cite{Tom2014,Tom2013,Tom2012,Tom2011} in energy and momentum due to the lack of translational symmetry is not apparent.
Interestingly, at $W_\mathrm{max}=20$meV [panel (b)] the momentum region without positive frequency reduces to a smaller one, indicating an increase of the length scale in which the ordering persists.
Most importantly, at $W_\mathrm{max}=30$meV [panel (c)] the magnon spectrum shows well-defined positive frequency in the entire momentum space, indicating that the proposed stripe ($\pi$,0) AFM order is a stable state of the system!
This confirms our proposal (cf. Fig.~\ref{fig1}) that by disturbing enough the carrier-induced long-range fluctuation via (chemical) disorder scattering, \textit{a strong electronic order can emerge from the previous failed order state.}


Figure.\ref{fig2} also shows a clear trend about the emergence of long-range order.
As the disorder increases, $\Delta q$ systematically decreases, reflecting the fact that the correlation is able to extend to a longer length scale $\sim 2\pi/\Delta q$.
Associated with it is the systematical reduction of the strength of the ``negative'' frequency (representing imaginary frequency) associated with the unstable magnon mode, indicating that the damaging long-range fluctuation systematically becomes weaker.
At the point when the strength is no longer able to negate the magnon frequency,  $\Delta q$ becomes zero and the correlation can extend to the system size and establish the long-range order.

\begin{figure}
	\setlength{\belowcaptionskip}{-0.5cm}
	\vspace{-0.1cm}
	\begin{center}
		\includegraphics[width=8cm]{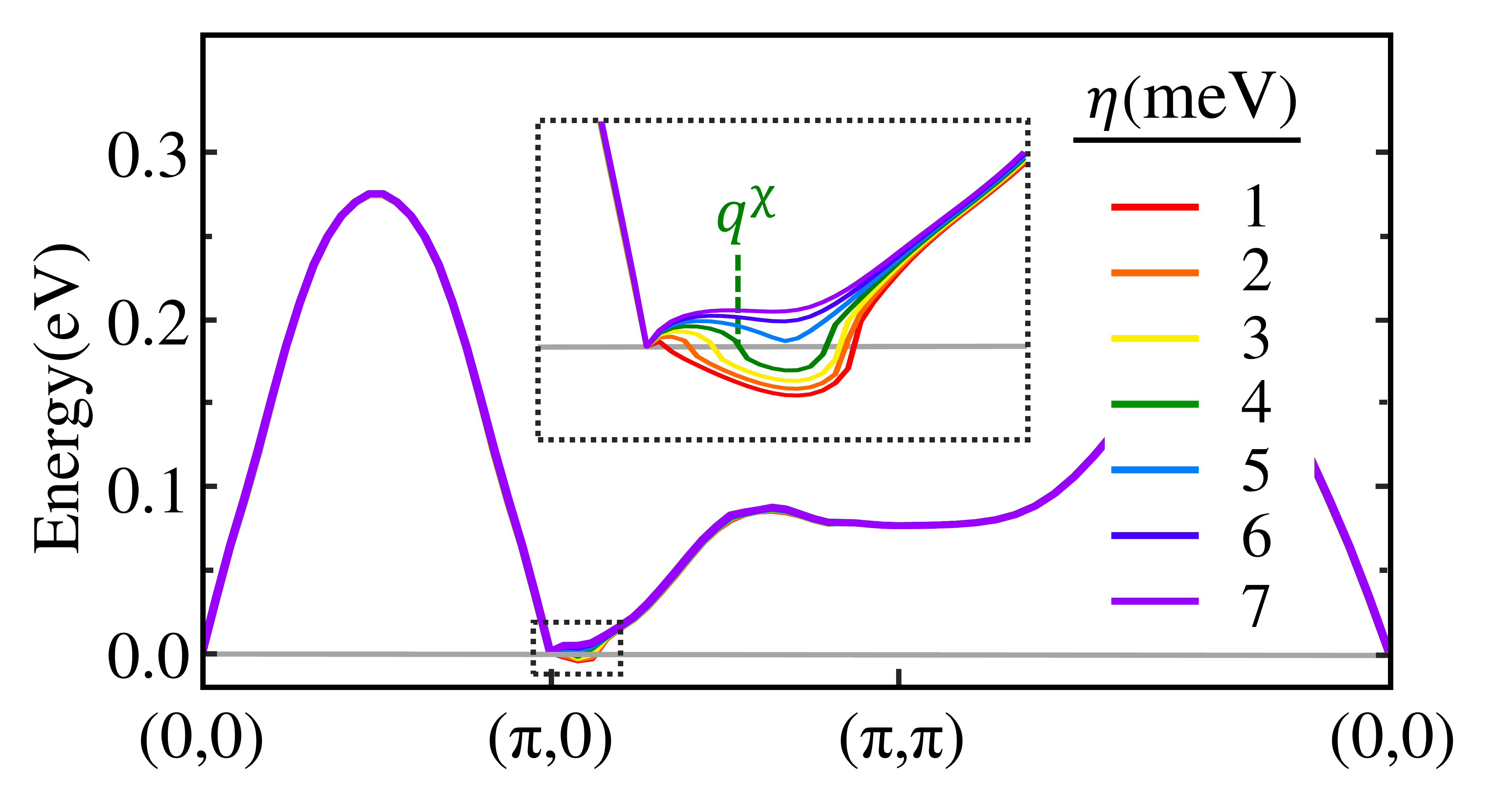}
	\caption{Magnon energy-momentum dispersion upon restricting the length scale of carrier propagation via different scattering rate $\eta$. With larger scattering rate (shorter length scale of coherent propagation), the region with imaginary frequency and the associated $\Delta q$ reduces.  Correspondingly, the momentum, $q^{\chi}$, at which the energy turns imaginary increases, indicating a longer $2\pi/\Delta q$ correlation resulting from a shorter length scale $2\pi/q^\chi$ of the carriers' damaging fluctuation.  By $\eta>$5meV, the excitation energy becomes fully positive, indicating the stability of the ($\pi$,0) AFM ordered state.}
	\label{fig3}
	\end{center}
\end{figure}

To gain further microscopic insight on how disorder scattering produces this unusual effect, notice that according to Eq.(\ref{J_phi_dis}), the main effect of the scattering is to induce a phase shift proportional to the path integral.
Therefore, one would expect the coherence of the renormalization of $\tilde{J}_{ii^\prime}$ to suffer systematically at longer range.
Particularly, beyond a characteristic \textit{length scale} that emerges when the random fluctuation of the phase reaches the order of 2$\pi$, the power-law tail of the carrier-induced fluctuation should no longer be effective.

To verify this simple intuition and to make a better connection with the underlying carrier dynamics, we investigate the effects of finite length scale of the carrier propagation on their quantum fluctuation (without the above disorder-induced phase shift) and in turn the influence on the magnon dispersion.
This is easily implemented by imposing a finite one-body scattering rate $\eta$ in Eq.(\ref{A2}) via $\delta=2\eta$.
Figure~\ref{fig3} summarizes the resulting magnon dispersion for $\eta=1$meV to 7meV.
Indeed, the strength of the imaginary frequency becomes weaker systematically as the scattering rate increases, and eventually all magnon frequency becomes positive at $\eta>5$meV, when the long-range order becomes a stable phase.
Notice that the momentum region without positive frequency and its associated $\Delta q$ scale reduces systematically, just like in the above cases with disorder.
As expected, in the aspect of allowing the correlation to grow in range and finally reach a long-range order, a reduction in the length scale of carrier propagation leads to a suppression of the long-range fluctuation similar to that caused by the disorder.

\begin{figure}
	 \setlength{\belowcaptionskip}{-0.5cm}
	\vspace{-0.1cm}
	\begin{center}
	    \includegraphics[width=8cm]{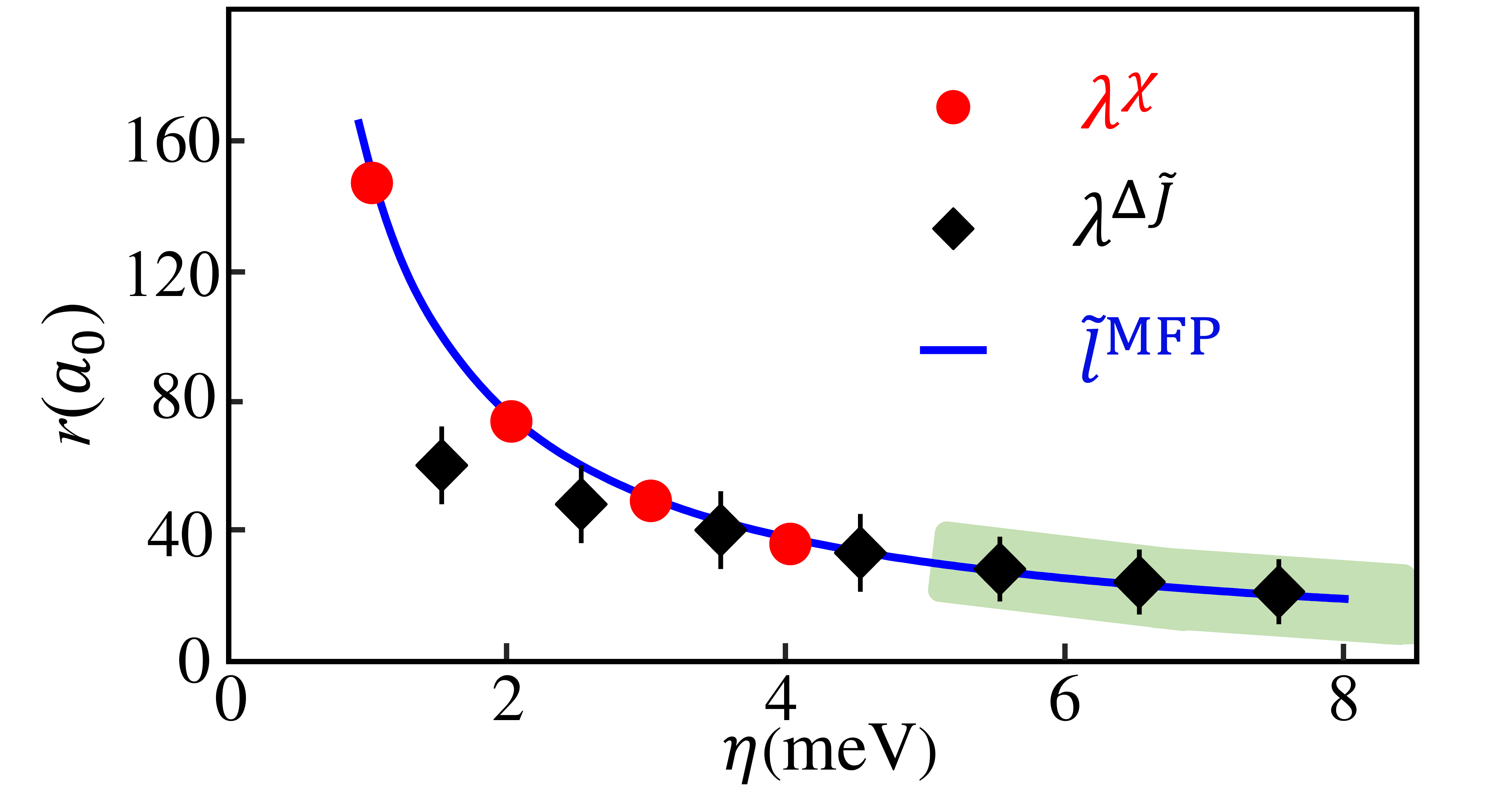}
	\caption{Correspondence of length scales in carrier-induced fluctuation and carrier propagation in unit of the lattice constant $a_0$.
    The length scale of carrier-induced damaging fluctuation, $\lambda^{\chi}=2\pi / q^{\chi}$, follows perfectly the scale of the mean-free path, $\tilde{l}^\mathrm{MFP} \propto v_{F}/\eta$.
    It also roughly corresponds to the length scale of the variation of the emerged long-range coupling, $\lambda^{\Delta \tilde{J}}$, defined as $\frac{\partial\widetilde{J}(r,\eta)}{\partial\eta}\bigg|_{r=\lambda^{\Delta \tilde{J}}}=0$.
    The light green region denotes the emergence of a stable ($\pi$,0) AFM ordered phase.
	}
	\label{fig4}
	\end{center}
\end{figure}

Figure.\ref{fig4} provides a more quantitative comparison between several relevant length scales in our results.
First, notice that in this length-scaled controlled picture, our results display a well-defined $q^\chi$ at which the magnon dispersion starts to become ``negative''.
It turns out that its corresponding length scale, $\lambda^\chi=2\pi/q^\chi$, follows perfectly the length scale of the mean-free path $\propto v_{F}/\eta \equiv \tilde{l}^\mathrm{MFP}$.
A similar consistency is also found in the length scale of the variation of the emerged long-range coupling~\cite{Supplementary}, $\lambda^{\Delta \tilde{J}}$, defined through $\frac{\partial\widetilde{J}(r,\eta)}{\partial\eta}\bigg|_{r=\lambda^{\Delta \tilde{J}}}=0$.
In essence, the limitation of the coherent length scale of the carrier-induced fluctuation, either through its phase-disordering or via scattering of carrier propagation itself, leads to a similar suppression of its effectiveness at long range, thereby allowing the correlation to extend to a longer range and eventually establish a long-range order (in the green region in Fig.\ref{fig4}).

While the above example concerns only the magnetic order, the underlying principles are generic to almost all symmetry breaking ordering since they mostly make use of only the general behavior of various mechanisms at long length scale.
For example, typical long-range order are driven by short-range many-body couplings that produce non-local correlation with a exponential decay at long range.
On the other hand, due to the discontinuity associated with the Fermi surface, the fermionic carrier-induced fluctuations usually have a long power-law tail that trumps the exponential decay of the above correlation.
This makes our proposed failed order more common than one might realize.
Indeed, in strongly correlated materials one often finds a rapid demise of finite-momentum long-range order upon enhancing metallicity, even though the correlations remain very strong at short range. 
As long as the damaging carrier-induced fluctuation only marginally overwhelms the ordering, our proposed mechanism would apply.
By suppressing via chemical disorder the carriers' ability to coherently interfere with the ordering at long range, the system has a chance to reveal its preferred long-range electronic order, in magnetic, orbital, charge, or other channels.

In short, to resolve the long-standing puzzle of the emergence of electronic order via the introduction of chemical disorder widely observed in strongly correlated metals, we propose a ``failed order'' scenario and verify it through a realistic spin-fermion model and a stability analysis based on linear response of the ordered state.
In essence, we propose that many of these strongly correlated metals are in a failed order state, in which the preferred order of the local moments is overwhelmed by carrier-induced long-range fluctuation.
The main effect of the chemical disorder is to efficiently reduce the coherent length scale of the damaging fluctuation and thereby allow the intrinsic long-range electronic order to emerge.
Our study demonstrates a typical example of the rich interplay between mechanisms of multiple length scales present in most strongly correlated metals, to which our proposed simple idea can be applied in general.

 

We thank Zhiqiang Mao, Vladimir Dobrosavljevi\ifmmode \acute{c}\else \'{c}\fi{}, Zi-Jian Lang, Athony Hegg, Fangyuan Gu, and Ruoshi Jiang for helpful discussions. 
This work is supported by the National Natural Science Foundation of China (NSFC) No. 11674220 and 12042507.

\bibliography{MainTex}

\clearpage
\pagebreak
\widetext
\begin{center}
\textbf{\large Supplementary materials: Chemical disorder induced electronic orders in correlated metals}
\end{center}
\begin{center}
\text{Jinning Hou, Yuting Tan, and Wei Ku}
\end{center}
\setcounter{equation}{0}
\setcounter{figure}{0}
\setcounter{table}{0}
\setcounter{page}{1}
\makeatletter
\renewcommand{\theequation}{S\arabic{equation}}
\renewcommand{\thefigure}{S\arabic{figure}}
\renewcommand{\bibnumfmt}[1]{[S#1]}
\renewcommand{\citenumfont}[1]{S#1}

This supplementary provides additional detailed information about our calculation that is based 
on well-established methods in the literature.

\section{Linear spin wave theory}
The spin Hamiltonian only contains collinear local moments
\begin{equation}
    H^{\mathrm{loc}}=\sum_{i\neq j}J_{ij}\mathbf{S}_{i} \cdot \mathbf{S}_{j},
\end{equation}
where the local moments $\mathbf{S}_{i}$ at site $i$ and $i^\prime$ couple via $J_{ii^{\prime}}$ ferromagneticlly ($J_{ii^{\prime}}<0$) or antiferromagneticlly ($J_{ii^{\prime}}>0$).
It is convenient to transform the collinear spin operators from local frame to lab frame via a spin rotation
\begin{equation}
    S_{i}^{x}=\tilde{S}_{i}^{x}, \ S_{i}^{y}=\kappa_{i} \tilde{S}_{i}^{y}, \ S_{i}^{z}=\kappa_{i} \tilde{S}_{i}^{z}, 
\end{equation}
where $\kappa_{i}=e^{i\mathbf{Q}\cdot \mathbf{r}_{i}} = \pm 1$.
The spin operators also can be expresses using raising operator $\tilde{S}^{+}_{i}$ and lowering operator $\tilde{S}_{i}^{-}$
\begin{equation}
    \tilde{S}_{i}^{+} = \tilde{S}_{i}^{x} + i \tilde{S}_{i}^{y}, \ 
    \tilde{S}_{i}^{-} = \tilde{S}_{i}^{x} - i \tilde{S}_{i}^{y}.
\end{equation}
In such rotated frame, all local moments point towards the ``up" direction.
Using the lowest-order Holstein-Primakoff (HP) transformation ($S$ is the magnitude of spin in the following discussion)
\begin{equation}
    \tilde{S}_{i}^{z}=S-a_{i}^{\dagger}a_{i},\ \tilde{S}_{i}^{+}=\sqrt{2S}a_{i}, \ \tilde{S}_{i}^{-}=\sqrt{2S}a_{i}^{\dagger},
\end{equation}
we can obtain the quadratic linear spin-wave Hamiltonian with bosonic creation operator $a_{i}^{\dagger}$and annihilation operator $a_{i}$ from local moments
\begin{equation}
\label{H_J_real_space}
   H^{\mathrm{\mathrm{loc}}} =\sum_{i} K_{i} a_{i}^{\dagger}a_{i}
    + \sum^\mathrm{FM}_{i\neq i'} J_{ii'}(a_{i}^{\dagger}a_{i'}+a_{i}a^{\dagger}_{i'})+\sum^\mathrm{AF}_{i\neq i''}J_{ii''}(a_{i}^{\dagger}a_{i''}^{\dagger}+a_{i}a_{i''}),
\end{equation}
where $K_{i} = 2\sum^\mathrm{AF}_{i''}J_{ii''}-2\sum^\mathrm{FM}_{i'}J_{ii'}$ ensures preservation of the Goldstone mode of the ordered system.
Here the summation is split into those between the parallel (FM) and anti-parallel (AF) pairs of spins.
The bosonic operators satisfy the commutation relations
\begin{equation}
\label{boson_commutation}
    [a_{i},a^{\dagger}_{i'}]=\delta_{ii'}, \quad [a_{i}^{\dagger},a_{i}^{\dagger}]=[a_{i},a_{i'}]=0.
\end{equation}

A simple one-band spin wave Hamiltonian represented in momentum $q$ space is
\begin{equation}
\label{H_loc_A0_B0}
    H^{\mathrm{loc}}=\sum_{q}J^{A}(q)(a_{q}^{\dagger}a_{q}+a_{-q}a^{\dagger}_{-q})+J^{B}(q)(a_{q}^{\dagger}a_{-q}^{\dagger}+a_{q}a_{-q}),
\end{equation}
where $J^{A}(q)$ is the coefficient after Fourier transformation of $a_{i}^{\dagger}a_{i'}$ ($i=i'$ or $i\neq i'$) and $J^{B}(q)$ is the coefficient after Fourier transformation of $a_{i}^{\dagger}a_{i''}^{\dagger}$.
The spin-wave dispersion is
\begin{equation}
    \omega(q)=\sqrt{(J^{A}(q))^{2}-(J^{B}(q))^{2}}.
\end{equation}

\section{Bogoliubov diagonalization of general quadratic bosonic Hamiltonian}
If the spin system does not have a simple translational symmetry and is difficult to diagonalize by hand,
we can use a general method~\cite{Tsallis1978} to diagonalize the quadratic bosonic Hamiltonian.
Considering a general quadratic bosonic Hamiltonian
\begin{equation}
\label{H_boson_quadratic}
    H=\sum_{ij}t_{ij}a_{i}^{\dagger}a_{j}+\tau_{ij}a^{\dagger}_{i}a_{j}^{\dagger}+\tau^{\star}_{ij}a_{i}a_{j},
\end{equation}
where $t_{ij}$ is the hopping parameter for a boson from index $j$ to $i$ and $\tau_{ij}$ is the parameter for creating two bosons with index $i$ and $j$.
If Hamiltonian $H$ is Hermitian, $t_{kj} = t_{ij}^{\star}$.
$\tau_{ij} = \tau_{ji}$ here.
We aim to get the diagonal Hamiltonian
\begin{equation}
    H=\sum_{i}\epsilon_{i}b^{\dagger}_{i}b_{i},
\end{equation}
where $\epsilon_{i}$ is the eigenvalues of index $i$ and bosonic operators $b_{i},b^{\dagger}_{i}$ satisfy the commutation relations Eq.(\ref{boson_commutation}).
The new operators are the linear combination of the previous operators
\begin{equation}
    b_{i}^{\dagger}=\sum_{j}a^{\dagger}_{j}T^{N}_{ji}+\sum_{j}a_{j}T^{A}_{ji}, \quad
    b_{i}=\sum_{j}a_{j}T^{N\star}_{ji}+\sum_{j}a_{j}^{\dagger}T^{A\star}_{ji},
\end{equation}
where $T^{N}_{ji}$ and $T^{A}_{ji}$ are the eigenvectors.
We can define a redundant and over complete basis:
\begin{equation}
\label{relation_A_B}
    A_{I}^{\dagger}=\left\{ 
\begin{array}{cc}
    a_{i}^{\dagger} & ;  I\in U \\
    a_{i} (i = I - \# \mathrm{\ of \ } i) &;  I\in D
\end{array} \right. , \quad
B_{I}^{\dagger} = \sum_{J}A_{J}^{\dagger}T_{JI},
\end{equation}
where $U$ and $D$ means the upper and down channel respectively.
$T_{JI}$ is the matrix of eigenvectors 
\begin{equation}
T_{JI} \longrightarrow\left(
\begin{array}{cc}
    T_{ji}^{N} & T^{A\star}_{ji} \\
    T_{ji}^{A} & T^{N\star}_{ji}
\end{array}\right).
\end{equation}
Since the bosonic satisfy the commutation relations Eq.(\ref{boson_commutation}), there is
\begin{equation}
\label{B_commu}
    [B_{I}, B^{\dagger}_{J}] = c_{I}\delta_{IJ}; \mathrm{\ where \ } c_{I}=\left\{\begin{array}{cc}
        1 &  ; I\in U\\
        -1 &  ; I\in D
    \end{array}\right. .
\end{equation}
Substituting Eq.(\ref{relation_A_B}) into Eq.(\ref{B_commu}), we can obtain the rule of orthonormalizing eigenvectors
\begin{equation}
    \sum_{I'J'}[A_{I'},A^{\dagger}_{J'}]T^{\star}_{I'I}T_{J'J}=\sum_{K}c_{K}T^{\star}_{KI}T_{KJ}=c_{I}\delta_{IJ}.
\end{equation}
Using the commutation relations Eq.(\ref{boson_commutation}), the result that the Hamiltonian Eq.(\ref{H_boson_quadratic}) commute with $B_{I}^{\dagger}$ is
\begin{equation}
\label{H_BI_commu}
    [H,B_{I}^{\dagger}]=c_{I}\epsilon_{I}B^{\dagger}_{I}=\sum_{J}A_{J}^{\dagger}T_{JI}c_{I}\epsilon_{I}.
\end{equation}
Eq.(\ref{H_BI_commu}) can be expressed using Eq.(\ref{relation_A_B}) as
\begin{equation}
\label{H_AI_commu}
    \sum_{K}[H,A_{K}^{\dagger}]T_{KI}=\sum_{k\in U}\left(\sum_{j}t_{jk}a_{j}^{\dagger} + \sum_{j}(\tau^{\star}_{kj}+\tau^{\star}_{jk})a_{j} \right) T_{KI} - \sum_{k\in D}\left(\sum_{j}t_{kj}a_{j} + \sum_{j}(\tau_{kj}+\tau_{jk})a_{j}^{\dagger} \right) T_{KI}.
\end{equation}
Combinding Eq.(\ref{H_BI_commu}) and Eq.(\ref{H_AI_commu}), we find
\begin{equation}
    T_{JI}c_{I}\epsilon_{I}=\sum_{K}M_{JK}T_{KI} \mathrm{\ ; \ where \ } M_{JK}=\left(\begin{array}{cc}
        t_{jk} &  -2\tau_{jk} \\
        2\tau_{jk}^{\star} & -t_{kj}
    \end{array}\right).
\end{equation}
The matrix $M$ is the non-Hermition matrix that we diagonalize and we can get the eigenvalues $\epsilon_{I}$ and corresponding eigenvectors.

\section{Integrating out the carriers}
Here, we derive the effective linear spin-wave Hamiltonian via integrating out the influence of itinerant carriers~\cite{lv2010,tam2015,Tan2019}. 
In general, the spin-fermion Hamiltonian contains local moments and itinerant carriers
\begin{equation}
\label{H_spin_fermi}
    H=H^{\mathrm{loc}}+H^{\mathrm{it}}+H^{\mathrm{H}},
\end{equation}
where $H^{\mathrm{loc}}$ is the spin Hamiltonian of local moments, $H^{\mathrm{it}}$ is the Hamiltonian of itinerant carriers and $H^{\mathrm{H}}$ describes the coupling between local moments and itinerant carriers.
For simplicity, we use one-band linear spin-wave Hamiltonian Eq.(\ref{H_loc_A0_B0}) and treat the Hund's coupling $J_{\mathrm{H}}$ between the itinerant and local degrees of freedom $\frac{J_{\mathrm{H}}}{2S}$ as perturbation (in unit of $S^2$)
\begin{equation}
    H^{\mathrm{H}}=-\frac{J_{\mathrm{H}}}{2S}\sum_{im\nu\nu^{\prime}}\mathbf{S}_{i}c^{\dagger}_{im\nu}\boldsymbol{\sigma}_{\nu\nu^{\prime}}c_{im\nu},
\end{equation}
where $\boldsymbol{\sigma}_{\nu\nu^{\prime}}$ are the Pauli matrices and $c^{\dagger}_{im\nu}$ represents creating an itinerant carrier at site $i$ of orbital $m$ with spin $\nu$.
Applying a canonical transformation to the Hamiltonian Eq.(\ref{H_spin_fermi})
\begin{equation}
    e^{\Delta}He^{-\Delta} = H + [\Delta,H] + \frac{1}{2}[\Delta,[\Delta,H]] + \dots
\end{equation}
results in the renormalized linear spin-wave Hamiltonian from its quadratic components
\begin{equation}
\label{H_SW_no_dis}
    H^{\mathrm{SW}}=H^{\mathrm{loc}}+<(H^{\mathrm{H}})^{2}+\frac{1}{2}[\Delta,(H^{\mathrm{H}})^{(1)}]>_{e} = \sum_{q}[\tilde{J}^A(q)(a_{q}^{\dagger}a_{q}+a_{-q}a_{-q}^{\dagger})+\tilde{J}^B(q)(a_{q}^{\dagger}a_{-q}^{\dagger}+a_{-q}a_{q})]
\end{equation}
up to the second order in $1/S$, where
\begin{equation}
    \tilde{J}^A(q)=J^A(q)+A(q), \quad \tilde{J}^B(q)=J^B(q)+B(q),
\end{equation}
and
\begin{equation}
\begin{split}
    A(q) = &\frac{J_{H}^{2}}{2S}\sum_{kll^{\prime}}\frac{f_{l}(k)-f_{l^{\prime}}(k+q)}{E_{l}(k)-E_{l^{\prime}}(k+q)} \times \left|\sum_{m}U^{l^{\prime}\star}_{m\downarrow}(k+q)U^{l}_{m\uparrow}(k)\right|^{2}\mathrm{,\ and}\\
    B(q) = &\frac{J_{H}^{2}}{2S}\sum_{kll^{\prime}}\frac{f_l(k)-f_{l^{\prime}}(k+q)}{E_{l}(k)-E_{l^{\prime}}(k+q)}\times \sum_{mm^{\prime}}U^{l^{\prime}\star}_{m\downarrow}(k+q)U^{l}_{m\uparrow}(k)U^{l\star}_{m^{\prime}\downarrow}(k)U^{l^{\prime}}_{m^{\prime} \uparrow}(k+q),
\end{split}
\end{equation}
where $E_l(k)$ denotes the eigenvalues with momentum $k$ and band index $l$ (that absorbs the spin index as well) and $U_{m\nu}^l(k)$ denotes the eigenvectors in the basis of orbital $m$ with spin $\nu=\uparrow$ or $\downarrow$.
$f_l(k)=\frac{1}{1+e^{\beta(E_l(k)-\mu)}}$ is the standard Fermi-Dirac distribution function for a given chemical potential $\mu$, and $S$ the effective magnitude of the local moments.
Note that $J^{A}(q)$ contains the a constant correction term $\frac{J_{H}}{2S}\sum_{kl}f_{l}(k)\sum_{m\nu}\nu|U^{l}_{m\nu}(k)|^{2} $ that is from the Hund's coupling. 

We also can get similar result using a dynamic method.
Using the perturbation theory with Green's function to integrate out the carrier channel, we can obtain the susceptibility in real part (see Eq.(3) in manuscript)
\begin{equation}
\begin{split}
    A(q) = &\frac{J_{H}^{2}}{2S}\sum_{kll^{\prime}}\frac{(f_{l}(k)-f_{l^{\prime}}(k+q))(E_l(k)-E_{l^{\prime}}(k+q))}{(E_{l}(k)-E_{l^{\prime}}(k+q))^2+(2\eta)^{2}} \times \left|\sum_{m}U^{l^{\prime}\star}_{m\downarrow}(k+q)U^{l}_{m\uparrow}(k)\right|^{2}\mathrm{,\ and}\\
    B(q) = &\frac{J_{H}^{2}}{2S}\sum_{kll^{\prime}}\frac{(f_l(k)-f_{l^{\prime}}(k+q))(E_{l}(k)-E_{l^{\prime}}(k+q))}{(E_{l}(k)-E_{l^{\prime}}(k+q))^2+(2\eta)^{2}}\times \sum_{mm^{\prime}}U^{l^{\prime}\star}_{m\downarrow}(k+q)U^{l}_{m\uparrow}(k)U^{l\star}_{m^{\prime}\downarrow}(k)U^{l^{\prime}}_{m^{\prime} \uparrow}(k+q),
\end{split}
\end{equation}
where the scattering rate $\eta$ of the carrier-induced fluctuation is dominated by the disorder effect, and is therefore set $0^+$ in the absence of disorder for a clearer comparison.
Diagonalization of the spin-wave Hamiltonian gives the spin-wave dispersion
\begin{equation}
    \omega(q)=\sqrt{(\tilde{J}^{A}(q))^{2}-(\tilde{J}^{B}(q))^{2}}.
\end{equation}
Using the renormalized linear spin wave Hamiltonian, we can obtain the renormalized Hamiltonian in real space in terms of Eq.(\ref{H_J_real_space}) via Fourier transformation.

\section{Weak disorder on the emerged long-range couplings}
Our goal is to investigate the effect of weak charge disorder on the spin channel. 
It is well-known~\cite{jagannathan1988,bulaevskii1986rkky,sobota2007rkky} that the main effect of disorder-induced scattering on the magnetic quantum fluctuation of itinerant carriers is to introduce incoherent phase shifts along its propagation without affecting its power-law  spatial decaying profile.
When the Fermi wavevector $k_F$ is well-defined, the oscillations with weak non-magnetic disorders can be expressed as~\cite{bulaevskii1986rkky} $J(r) \cos{(2k_{F}r+\phi_{r})}$, where $J(r)$ is the magnitude with power-law decaying, $r$ is the distance from site $i$ to site $i^{\prime}$ in real space.
In discrete lattice, Eq.(\ref{H_SW_no_dis}) can be transformed into Eq.(\ref{H_J_real_space}) via Fourier transformation, $J_{ii^{\prime}}\cos{(2\mathbf{k_{F}}(\mathbf{x}_i-\mathbf{x}_{i^{\prime}}))}$.
In realistic systems, however, the Fermi surface is typically not perfectly nested and thus the oscillation in $J_{ii^\prime}$ is not with a fixed $2k_F$ period, but instead it displays a rather complicated pattern.
We therefore approximate the disorder-induced phase shift via
\begin{equation}
    \tilde{J}_{ii^{\prime}} \longrightarrow \tilde{J}_{ii^{\prime}}\cos{\phi_{ii^{\prime}}},
\end{equation}
where $\tilde{J}_{ii^{\prime}}$ contains the power-law decaying term and oscillating term.
The disorder-dependent phase is
\begin{equation}
    \phi_{ii^\prime} = \frac{2}{\hbar v_{F}} \int_{\mathbf{x}_{i}}^{\mathbf{x}_i^\prime} ds W(\textbf{x}),
\end{equation}
where $W(\textbf{x})$ denotes the strength of the spatial disorder randomly sampled from a uniform distribution between 0 and $W_\mathrm{max}$ and the integration is along a straight path from position $\textbf{x}_{i}$ of site $i$ to position $\textbf{x}_i^\prime$ of site $i^\prime$.
Here $v_F$ is the Fermi velocity and $\hbar$ is 1 in the atomic unit.
Then we discretize the disorder-dependent phase factor from a continuum space to a discrete lattice.

\begin{figure}
	\begin{center}
		\includegraphics[width=10cm]{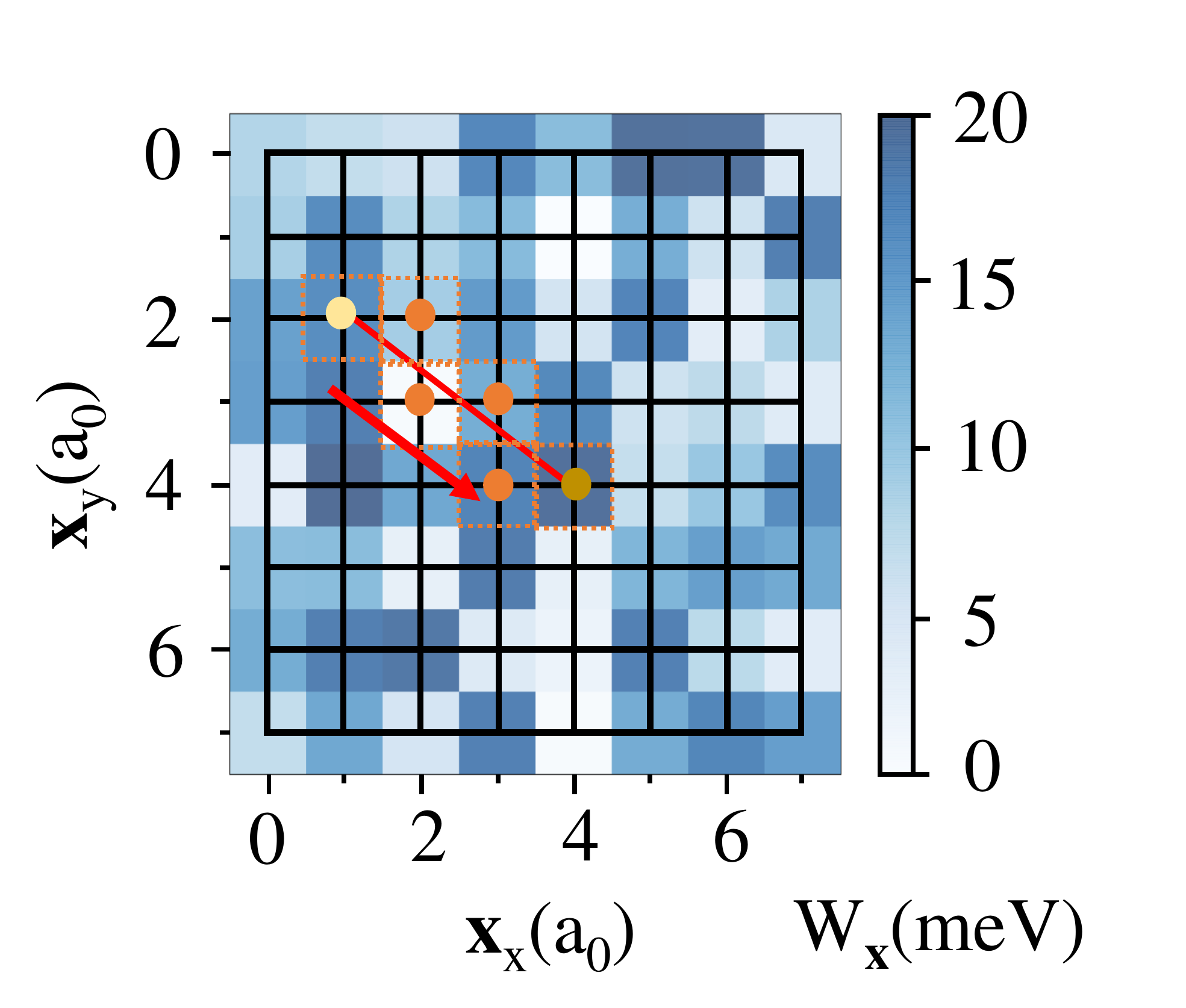}
	\caption{An example path integral of the disorder-dependent phase factor in a discrete 8$\times$8 lattice from light yellow site (1,2) to dark yellow site (4,4).  Along the straight path, it would pass a set of squares with different depths of blue color which represent the strength of potential energy $W(\bf{x}_{i})$. }
	\label{Sfig1}
	\end{center}
\end{figure}

\begin{figure}
	\begin{center}
		\includegraphics[width=19cm]{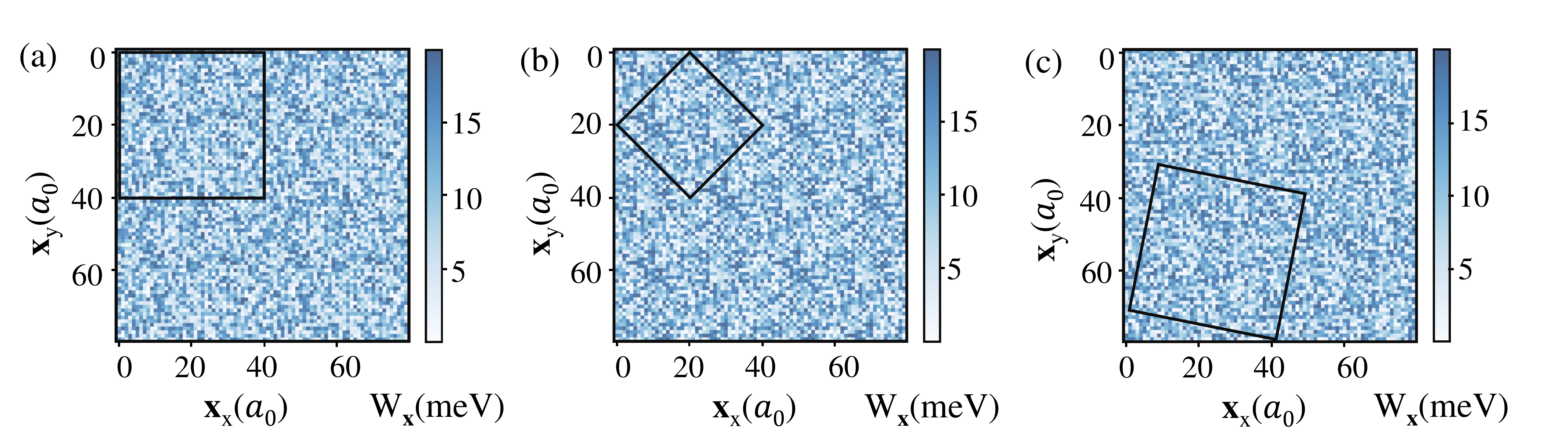}
	\caption{Examples of disorder configurations with the maximum value $W_{\mathrm{max}}=20$meV in a 80$\times$80 lattice. The black dotted squares show disordered patterns which have different sizes and orientations.}
	\label{Sfig3}
	\end{center}
\end{figure}

An example of discretizing the phase factor is shown as Fig.\ref{Sfig1}.
Every site at the lattice have different random potential energy from zero to a maximum potential energy $W_{\mathrm{max}}$.
We treat the potential energy $W_{\textbf{x}_{i}}$ dominating a square range around the site $\textbf{x}_{i}$.
The total phase factor is the summation of $W_{\textbf{x}_{i}}\times ds$ from $\textbf{x}_{i}$ to $\textbf{x}_{i^{\prime}}$.
$ds$ is the length in the square range around the disorder site.
We generate random potential $W_{\textbf{x}_{i}}$ in lattice with different sizes and orientations in the range of $(0, W_{\mathrm{max}})$.
Three kinds of disorder configurations as examples as shown in Fig.\ref{Sfig3}.
The Fermi velocity is estimated via the derivation of the Hamiltonian along $y$ direction around Fermi energy.

\section{Length scale of variation of the emerged long-range coupling}
We get the renormalized spin wave Hamiltonian by integrating out the itinerant carriers with different scattering rate $\eta$:
\begin{equation}
    H^{\mathrm{SW}} = \sum_{q}\tilde{J}^A(q)(a^{\dagger}_{q}a_{q}+a_{-q}a^{\dagger}_{-q}) + \tilde{J}^B(q)(a^{\dagger}_{q}a^{\dagger}_{-q}+a_{-q}a_{q}).
\end{equation}
The Hamiltonian in momentum space can be transformed into real space Eq.(\ref{H_J_real_space}) with long-range couplings.
Since our system is mainly unstable along $q_{y}-$direction, we summate the contributions of $\tilde{J}_{ij}$ along $x-$direction and obtain the fluctuating decaying couplings along $y-$direction at long distance.

\begin{figure}
	\begin{center}
		\includegraphics[width=12cm]{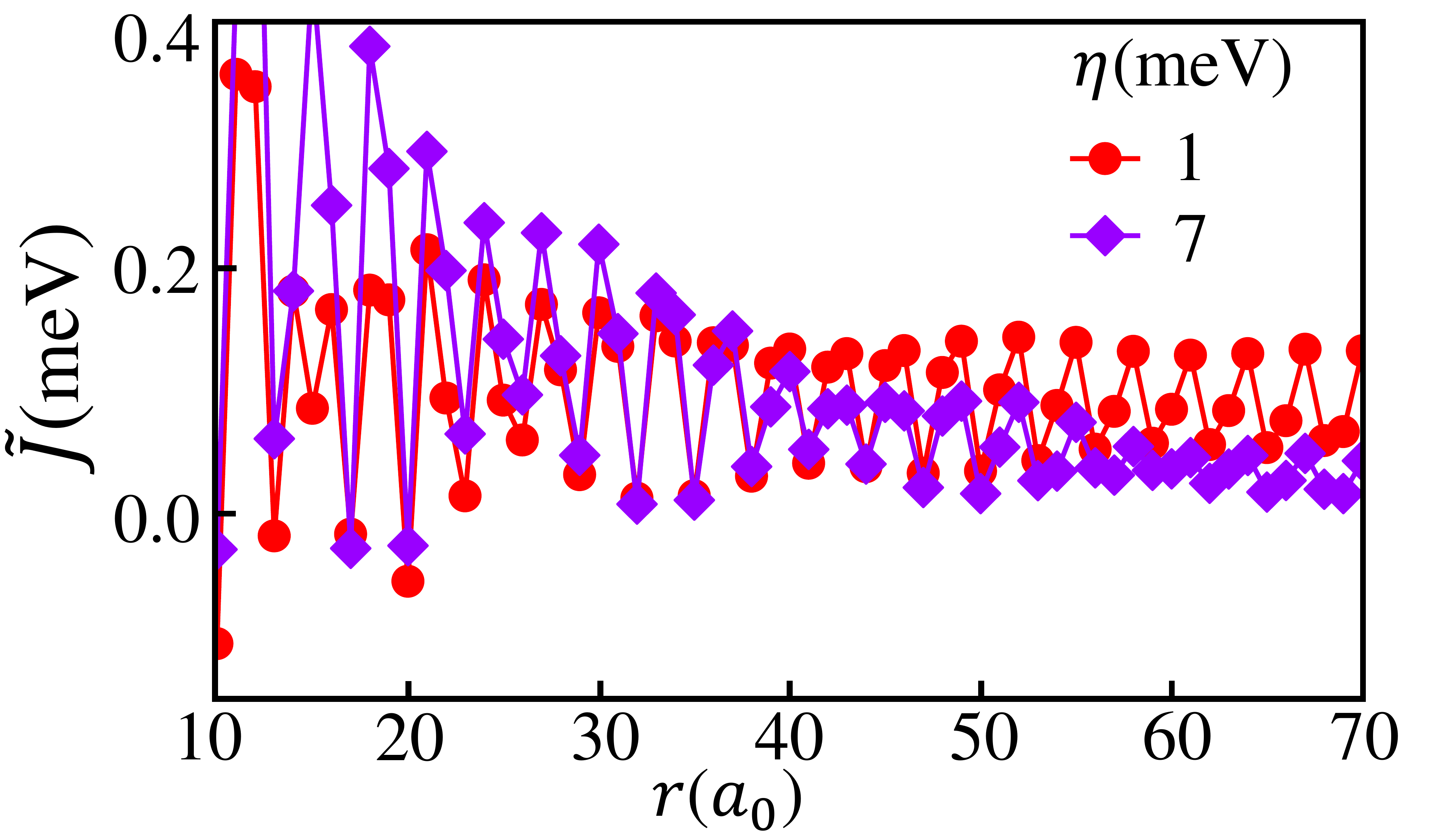}
	\caption{Renormalized spin-spin interaction $\tilde{J}$ along $y$ direction at the range from 10$a_{0}$ to 70$a_{0}$ with scattering rate 1meV and 7meV. $\tilde{J}$ is oscillating decaying. The two different lines across at this range that shows the long-range coupling is suppressed when disorder is stronger.}
	\label{Sfig2}
	\end{center}
\end{figure}

Figure.~\ref{Sfig2} shows the renormalized spin-spin interaction $\tilde{J}$ at long distance with different $\eta=1$meV and $\eta=7$meV as example.
The coupling is suppressed at long range and enhanced at short range with increasing damping energy.
Since lines with different $\eta$ would cross with each other, we define a distance $\lambda^{\Delta \tilde{J}}$ defined as $\frac{\partial\widetilde{J}(r,\eta)}{\partial\eta}\bigg|_{r=\lambda^{\Delta \tilde{J}}}=0$ and estimate the range.

\end{document}